\documentclass[twocolumn,showpacs,preprintnumbers,amsmath,amssymb,prb,superscriptaddress]{revtex4}

\usepackage{graphicx}
\usepackage{dcolumn}
\usepackage{bm}
\usepackage{natbib}

\begin{document}

\preprint{PREPRINT}

\title{Field emission mechanisms of graphitic nanostructures}%

\author{Masaaki Araidai}
\email{j1204701@ed.kagu.tus.ac.jp}
\affiliation{
Department of Physics, Faculty of Science, Tokyo University of Science,
1-3 Kagurazaka, Shinjuku-ku, Tokyo 162-8601, Japan
}
\affiliation{
CREST, Japan Science and Technology Agency, 4-1-8 Honcho Kawaguchi, Saitama 332-0012, Japan
}

\author{Yasuhiro Nakamura}
\altaffiliation[]
{Present address: Department of Materials Engineering, The University of Tokyo,
7-3-1 Hongo, Bunkyo-ku, Tokyo 113-8656, Japan}
\affiliation{
Department of Physics, Faculty of Science, Tokyo University of Science,
1-3 Kagurazaka, Shinjuku-ku, Tokyo 162-8601, Japan
}

\author{Kazuyuki Watanabe}
\email{kazuyuki@rs.kagu.tus.ac.jp}
\affiliation{
Department of Physics, Faculty of Science, Tokyo University of Science,
1-3 Kagurazaka, Shinjuku-ku, Tokyo 162-8601, Japan
}
\affiliation{
CREST, Japan Science and Technology Agency, 4-1-8 Honcho Kawaguchi, Saitama 332-0012, Japan
}

\date{\today}

\begin{abstract}
Field emission (FE) and the electronic-states origin of graphitic nanostructures 
were investigated by first-principles calculations based on time-dependent density-functional theory.
We find that the FE current from graphitic ribbons changes remarkably depending on
the hydrogen termination and the direction of the applied electric field.
Also, the FE current from graphene sheets shows a dramatic increase around vacancy defects.
We verified, through the analysis of local electronic structures and energy distributions of
emitted electrons, that the dangling-bond (or $\sigma$) character is responsible for these results
and governs the nature of the FE of graphitic nanostructures.
\end{abstract}



\maketitle

\section{Introduction}
\label{intro}

Recent advances in nanotechnology have made
possible the fabrication of a wide variety of nanometer-scale devices using graphitic nanostructures,
such as fullerenes,~\cite{Fullerene} carbon nanotubes,~\cite{CNT}
and other carbon-based materials.~\cite{Rodriguez,Wu}
Among the various applications for these nanostructures,
electron field emitters made from carbon allotropes show significant promise
for novel electronic devices,~\cite{Choi,Yahachi,Geis,Cui}
because they can maintain stable forms under extremely high field-emission (FE) current densities
owing to tight covalent-bonds.
Chen \textit{et al.} performed an FE experiment using graphite platelet nanofibers (GPNs),~\cite{Chen}
with several thousand graphitic ribbons stacked together like a deck of cards.~\cite{Rodriguez}
Wu \textit{et al.} carried out FE measurements of carbon nanowalls (CNWs),
which are nano-graphitized sheets grown perpendicularly on substrates,~\cite{Wu,Wu2}
and pointed out that CNWs are good candidates for nanoscale field emitters as well as carbon nanotubes.~\cite{Wu2}
For the development of efficient field emitters,
examination of FE properties of graphitic nanostructures is
interesting and meaningful both from the viewpoint of fundamental science
as well as technological applications of the field emitters.

Over the past few years, interesting phenomena in graphitic ribbons have been predicted by
theoretical studies.~\cite{Fujita,Nakada,Wakabayashi,Miyamoto,Kawai}
According to theoretical results based on the tight-binding model
for the $\pi$ electrons,~\cite{Fujita,Nakada,Wakabayashi}
zigzag graphitic ribbons are always metallic, while armchair graphitic ribbons are
either metallic or insulating depending on the widths of the ribbons.
The study also pointed out that zigzag ribbons exhibit specific electronic states,
highly localized at the ribbon edges (edge states).
This was supported by first-principles calculations~\cite{Miyamoto,Kawai}
based on density-functional theory (DFT).~\cite{Kohn}
For the edge state, magnetic,~\cite{Fujita,Wakabayashi} optical,~\cite{Lin}
and thermal~\cite{Yamamoto} properties of graphitic ribbons have also been studied theoretically.
However, thus far, there have been very few experimental studies of graphitic ribbons.~\cite{Terai,Cancado}

Recently, the FE of graphitic ribbons was investigated by Tada and Watanabe 
using the time-dependent density-functional theory (TD-DFT).~\cite{Tada_PRL}
They found that the dangling-bond states and not the edge states contribute primarily to the FE.
They also found that not only evaluation of work functions
but also knowledge of local electronic properties, the $\sigma$- or $\pi$-bonding states,
are prerequisite for understanding the microscopic mechanisms of FE properties of covalently-bonded nanostructures.
Tada and Watanabe's results cannot be derived from the conventional Fowler-Nordheim theory~\cite{FN}
for the FE from flat jellium surfaces with the free-electron approximation.
With the ultimate goal of complete understanding of microscopic mechanisms of the FE of graphitic nanostructures,
in this study, we have investigated the effects of the electric-field direction and
vacancy defects on the FE characteristics of graphitic nanostructures by using TD-DFT calculations.

The outline of this paper is as follows.
In Sec.~II, the computational methods and models used are described.
In Sec.~III, we present numerical results for the effects of hydrogen (H) termination
and the direction of applied electric field on the FE of graphitic ribbons.
We also present results for the effect of vacancy defects on the FE of graphene sheets.
We conclude with Sec.~IV.

\section{Methods and Models}
\label{method}

We carried out conventional DFT~\cite{Kohn} and TD-DFT~\cite{Runge} calculations
to investigate the electronic states and electron emission properties of graphitic ribbons and graphene sheets.
TD-DFT~\cite{Runge} has been successfully applied to various 
non-equilibrium electron dynamics phenomena,~\cite{Yabana,Sugino}
including FE of carbon nanotubes,~\cite{Han_CNT} graphitic ribbons,~\cite{Tada_PRL}
covalently-bonded atomic wires,~\cite{Han_chain,Araidai2} and diamond surfaces.~\cite{Araidai,Araidai3}
We employed the norm-conversing pseudopotentials of NCPS97~\cite{Kobayashi}
based on the Troullier-Martins algorithm~\cite{Troullier}
and the generalized gradient approximation by Perdew \textit{et al}.
for the exchange-correlation potential.~\cite{Perdew}
The block Davidson algorithm was adopted to diagonalize the Kohn-Sham Hamiltonian matrix.
The electronic wave functions were expanded in terms of a plane-wave basis set.
In the TD-DFT calculations, the expansion coefficients become time-dependent quantities.

First, we determined the ground states of atomic and electronic structures
by using conventional DFT in a zero electric field.
Next, we applied an electric field of 10 V/nm
and calculated the time evolution of the wave functions by applying the TD-DFT scheme.
Then, wave functions with higher energies start to contribute to FE.
The KS Hamiltonian is updated by a new electron density at each time step,
and a part of the dynamical screening effect and some electron correlation effects
are automatically taken into account in this calculations.
It is noted that the simulation time should not exceed a certain time
because the emitted electrons bounce back from the unit-cell boundaries.
However, within a simulation time $t$ $\leq$ 150 a.u. (3.6 fs), such reflection does not occur.
In this calculation of the time evolution of wave functions,
we have adopted the seventh-order Taylor expansion method.~\cite{Yabana}
The Taylor expansion method cannot guarantee the unitarity condition,
although this method has computational simplicity.
The Suzuki-Trotter type split-operator method~\cite{Suzuki}
would be more accurate and reliable for long time simulations,~\cite{Sugino}
because this method ensures the unitarity condition for the time evolution.
The Taylor expansion method, however, guarantees
the normalization condition for the electron number
with an accuracy of $10^{-6}$ for our simulation time.
Moreover, the computational time for the Taylor expansion method is much shorter
than that for the Suzuki-Trotter expansion method.
Therefore, we have employed the Taylor expansion method in this study.
Finally, we counted the number of electrons tunneling into a large vacuum region
by summing the squares of the coefficients at each time step
and obtained a curve of the electron number as a function of time.
The value of the FE current was evaluated from the linear slope in the curve.

The systems we investigated in this study are zigzag graphitic ribbons with and without H termination
and graphene sheets with and without a vacancy defect, as shown in Fig.~\ref{fig:model}.
Since the effect of edge states on the FE is interesting, we chose the zigzag ribbons in the present calculations.
The $x$ and $y$ axes are defined as shown in Fig.~\ref{fig:model}.
The unit cell sizes for the graphitic ribbons are 79.8 {\AA} $\times$ 2.49 {\AA} $\times$ 4.99 {\AA}
and 14.8 {\AA} $\times$ 2.49 {\AA} $\times$ 88.6 {\AA},
when electric fields are applied parallel ($x$) and perpendicular ($z$) to the ribbons, respectively.
The widths of the graphitic ribbon in Fig.~\ref{fig:model}(a) and (b) are 9.12 {\AA} and 7.10 {\AA}, respectively.
The size of the unit cell for the graphene sheets is 8.52 {\AA} $\times$ 7.38 {\AA} $\times$ 85.2 {\AA}.
The cutoff energies for the calculations for graphitic ribbons
and graphene sheets are 45 Ry and 34 Ry, respectively.
We chose 10 $k_y$ points in the first Brillouin zone for the graphitic ribbons
and 16 points in the $k_x$-$k_y$ plane in the first Brillouin zone for the graphene sheets
for the determination of the ground-state electronic and atomic structures.
The atomic structure around the vacancy defect of the graphene sheet was
noticeably deformed, as seen in Fig.~\ref{fig:model}(d).

\section{Results and Discussions}
\label{results}

\subsection{Direction of electric field}

We begin with a brief review of a prior study~\cite{Tada_PRL} on the FE
of zigzag ribbons with and without H termination
in a parallel electric field, $E_{\!/\!\!/}$, along the $x$ direction in Fig.~\ref{fig:model}.
This previous study clarified the effects of H-termination on the FE properties of the zigzag ribbons 
using local electronic structures.
The results of energy distributions of the emission currents
and the electronic states responsible for the peaks are shown by broken lines in Fig.~\ref{fig:dist}.
For the H-terminated ribbon (Fig.~\ref{fig:dist}(a)),
there are two peaks originating from $\sigma$ states in the energy distribution.
The electronic distribution of emitted electrons resulting in the higher peak
is shown in the upper right panel in Fig.~\ref{fig:dist}(a).
The edge state ($\pi$ state) is not a contributor to the FE of the H-terminated ribbon,
even though the edge state is near the Fermi level for the H-terminated ribbon.
On the other hand, for the clean ribbon (Fig.~\ref{fig:dist}(b)),
a prominent peak appears in the energy distribution.
The electronic-states origin of the sharp peak can be gleaned from
the upper right panel in Fig.~\ref{fig:dist}(b) to be the dangling-bond (DB) state.
The reason why electrons are emitted from the $\sigma$ state (H-terminated) or the DB state (clean),
and not from the $\pi$ states, can be elucidated from the directional angle
between the local electronic distributions and the applied field, $E_{\!/\!\!/}$.
Typical electronic orbitals, $\pi$, $\sigma$ and DB around an edge of the clean ribbon
are schematically shown in Fig.~\ref{fig:edge}.
As seen in Fig.~\ref{fig:edge},
$\sigma$ or DB are parallel to the parallel field, $E_{\!/\!\!/}$, and tend to respond to the field.

To verify that the electronic states contribute to the FE
as long as the electronic orbitals protrude along the direction of the electric field,
we have investigated whether $\pi$ orbitals react to an electric field
($E_\perp$, $z$ axis in Fig.~\ref{fig:model}) that is perpendicular to the ribbon sheet.

We now describe the results for the FE of the zigzag ribbon in the perpendicular field, $E_\perp$.
The energy distributions of the FE current and the corresponding electronic distributions
are given by solid lines in Fig.~\ref{fig:dist}.
For the H-terminated ribbon (Fig.~\ref{fig:dist}(a)),
FE currents are emitted from the $\pi$ states at the edge, as expected,
although no prominent peaks appear, compared to the energy distribution curve for the parallel field,
$E_{\!/\!\!/}$ (broken line in Fig.~\ref{fig:dist}(a)).
For the clean ribbon (Fig.~\ref{fig:dist}(b)), however,
a sharp peak originating from the DB state appears, similar to the case for the parallel field, $E_{\!/\!\!/}$.
Negligibly small contributions of the $\pi$ states to the total FE current are found in the energy distributions.
The direction of electronic distribution of the DB state
in the upper right panel in Fig.~\ref{fig:dist}(b) is \textit{not} parallel to the perpendicular field, $E_\perp$.
It follows from this result that the DB state contributes to the FE
even when the direction of the electronic distribution
is \textit{not} parallel to the direction of the electric field.

Here, it is important to clarify why $\pi$ states cannot be the main source of the FE
even when $\pi$ orbitals extend along the applied electric field, $E_\perp$ (See Fig.~\ref{fig:edge}).
Fig.~\ref{fig:field_profile} shows the equipotential surface of the clean ribbon on the plane perpendicular to the ribbon sheet
(inset) and electrostatic potential energy curves
along the direction of the perpendicular field, $E_\perp$.
It becomes apparent from the equipotential surface that the electric field becomes negligibly small
due to the screening effect in the center of the ribbon.
Therefore, electrons are not emitted from the center region of the ribbon.
On the other hand, the electric field is enhanced at the edges of the ribbon,
as seen from the equipotential surface,
and thus electrons are emitted mainly from the edges.
We focus on the three electrostatic potential energy curves drawn along the lines
corresponding to the three positions in the inset of Fig.~\ref{fig:field_profile}.
The slope of the potential curve around the origin of the horizontal axis
is the strength of the local electric field around the edge atoms
along the direction of the perpendicular field, $E_\perp$.
The direction of the local field of curve $1$, where the DB orbital exists,
is parallel to the applied field, $E_\perp$.
However, the direction of the field for curve $3$, where $\pi$ orbitals protrude,
becomes \textit{opposite} to that of the applied field.
Consequently, electrons cannot be emitted from the $\pi$ orbitals
even when the $\pi$ orbitals are parallel to the applied field.
The electrostatic potential energy curves of the H-terminated ribbon
also show similar features as the clean ribbon.

\begin{table}[b]
  \caption{
    FE currents $\mu$A/(unit cell) from two type of ribbons (H-termination or clean)
    in electric fields (10 V/nm) along two directions ($E_{\!/\!\!/}$ or $E_\perp$).
  }
  \begin{ruledtabular}
    \begin{tabular}{cccc}
      Type of termination & $E_{\!/\!\!/}$ &  $E_\perp$  \\ \hline
              Clean       &       0.37     &     0.10    \\
          H-termination   &       0.14     &     0.06    \\
    \end{tabular}
  \end{ruledtabular}
  \label{tab:table1}
\end{table}

Finally, we can summarize the FE properties of zigzag graphitic ribbons.
We list the values of the FE currents from the two type of ribbons, H-termination and clean,
in electric fields along two directions ($E_{\!/\!\!/}$ or $E_\perp$) in Table~\ref{tab:table1}.
The DB orbital is the main source of the FE from the clean ribbon
in both electric field directions.
The DB state disappears upon H-termination
and thus the $\sigma$ orbitals and the $\pi$ orbitals become emitting sources
in the parallel and perpendicular electric fields,
although the contributions of the $\pi$ orbitals are negligible.

\subsection{Vacancy defects}

Having clarified the important role of the DB state in the FE from the graphitic ribbons,
we further explored the FE property of the graphene sheet with an atomic vacancy (Fig.~\ref{fig:model}(d)),
because the DB states are generated in the vacancy.
The graphene sheets investigated are shown in Fig.~\ref{fig:model}(c) and (d).
The electric field was applied in a direction perpendicular ($E_\perp$) to the sheet.

We investigated the FE current-density images of the graphene sheets
to clarify the role of the vacancy on FE characteristics.
Figure~\ref{fig:current_dist} shows the FE images for the graphene sheet (a) without and (b) with vacancy defects.
The plane on which the FE current distribution is plotted is 4.9 {\AA} above the graphene sheet.
The FE current density around the vacancy is about 22 times as large as that from the other regions.
The total FE currents obtained from the graphene sheet without and with vacancy defects
are 0.054 and 0.106 $\mu$A/(unit cell), respectively.
The FE images thus obtained reflect a dramatic change in the electronic distribution around the vacancy defect
due to the removal of a carbon atom.
It becomes apparent from the FE images that vacancy defects significantly enhance
the FE current density of the graphene sheet due to the appearance of the DB states.
It should be noted that the FE image in Fig.~\ref{fig:current_dist}(b),
which is taken 4.9 {\AA} above the graphene sheet, does not necessarily indicate atomic-scale resolution FEM
because the electron beams from each vacancy site interfere with each other
before reaching the screen that is far from the emitter.

\section{Conclusions}
\label{conclusions}

The FE and the electronic-states origins of graphitic nanostructures
were investigated by first-principles calculations based on TD-DFT.
We found that the character of the local electronic states responsible for FE changes
and thus the FE current varies, depending on the conditions of hydrogen termination
and the direction of the electric field.
The DB states are predominant sources for the FE
because the electronic orbitals tend to appear at the edges
and protrude along the direction of the electric field in vacuum.
High FE currents from the edges of the graphene sheets of CNWs,~\cite{Wu2}
which are in parallel to the electric field, have been observed.
Thus, the present study can provide a theoretical interpretation for the features observed in the experiment.
We naturally expect that a similar property will be observed in the FE of GPNs.~\cite{Chen}
It is interesting to note that the direction of electronic-orbitals has been found to
remarkably influence the ionization rate of atoms in strong laser fields.~\cite{Ullrich}

Substantially enhanced emission current from the atomic vacancy sites in the graphene sheet
confirms the essential contribution of the DB states to the FE from graphitic nanostructures.
The $\pi$ states, however, contribute minimally to the FE
even when the electronic orbitals project along the direction of the electric field
because the electric-field strength tends to decrease substantially in the region of the $\pi$ orbitals,
\textit{i.e.}, owing to a screening effect.

These findings are remarkable reflecting the covalent-bond character.
The results emphasize the need for first-principles studies taking into account 
the geometric and electronic structures of field emitters
towards better understanding of FE mechanisms, and for the design of nanoscale graphitic field emitters.

\begin{acknowledgements}
The authors thank Taiju Fujiki for help with numerical calculations.
The present study was partly supported by the Ministry of Education, Sports, Culture,
Science and Technology of Japan through the Grant-in-Aid No.~15607018.
One of the authors (M.~A.) wishes to acknowledge the support of
the JSPS (Japan Society for the Promotion of Science) Research Fellowship for Young Scientists.
Part of the numerical calculations were performed on the Hitachi SR8000s at
the Supercomputer Center, Institute for Solid State Physics, University of Tokyo.
\end{acknowledgements}

\begin{figure}
  \caption{
           Top views of graphitic ribbons (a) with and (b) without hydrogen (H) termination
           and graphene sheets (c) without and (d) with a vacancy defect in the unit cell.
           Gray circles are carbon atoms and the small black dots in (a) are H atoms.
           Dashed straight lines denote boundaries of the unit cells
           and a dashed circle in (d) denotes a vacancy defect.
          }
  \label{fig:model}
  \caption{Energy distribution of the FE current for (a) the H-terminated ribbon and (b) the clean ribbon.
           The vacuum level is chosen for the origin of the energy.
           On the right side of each panel, electronic distributions causing the peaks
           are shown by blue clouds with carbon atoms (white spheres) and H atoms (white dots).
           The broken and solid lines denote the results in a parallel ($E_{\!/\!\!/}$)
           and perpendicular ($E_\perp$) electric field, respectively.
          }
  \label{fig:dist}
  \caption{
           Schematic diagram of electron orbitals (gray clouds)
           of a clean graphitic ribbon around the right edge (side view).
           $E_{\!/\!\!/}$ and $E_\perp$ denote the applied electric fields
           parallel and perpendicular to the ribbon, respectively.
           DB represents a dangling-bond.
          }
  \label{fig:edge}
  \caption{
           The electrostatic potential energy as a function of the position from the ribbon sheet.
           The plane on which the ribbon lies is chosen to be the origin of the horizontal axis.
           The inset shows the equipotential surface of the clean ribbon from a lateral view
           in the perpendicular field, $E_\perp$.
           Electrons are emitted from the ribbon into the upper region.
           The lines 1, 2 and 3 are 1.2 {\AA}, 0.6 {\AA} and 0 {\AA}
           away from the right edge atom of the ribbon.
          }
  \label{fig:field_profile}
  \caption{FE images of the graphene (a) without and (b) with a vacancy defect.
           The distance between the FE image plane and the graphene is 4.9 {\AA}.
           The FE current density around the vacancy is about 22 times larger than that from the other regions.
          }
  \label{fig:current_dist}
\end{figure}

\end{document}